\documentstyle[12pt,aps,prl,epsfig,aps,epsf,floats]{revtex}
\tighten

\newcommand{\bt}{\begin{tabular}}
\newcommand{\et}{\end{tabular}}
\newcommand{\be}{\begin{equation}}
\newcommand{\ee}{\end{equation}}
\newcommand{\bq}{\begin{eqnarray}}
\newcommand{\eq}{\end{eqnarray}}

\begin{document}
\title{
Experimental Search for Molecular-Nuclear Transitions in Water\thanks{%
LANL e-print nucl-ex/0001005}}
\author
{V.\,B.\,Belyaev$^{\dagger}$\thanks{E-mail: belyaev@thsun1.jinr.ru}
\and M.\,B.\,Miller$^{\ddagger}$
\and A.\,K.\,Motovilov$^{\dagger}$
\and \mbox{A.\,V.\,Sermyagin$^{\ddagger}$}
\and  I.\,V.\,Kuznetzov$^{\dagger}$
\and  Yu.\,G.\,Sobolev$^{\dagger}$
\and  A.\,A.\,Smolnikov$^{\dagger}$$^{\S}$
\and  A.\,A.\,Klimenko$^{\dagger}$$^{\S}$
\and  S.\,B.\,Osetrov$^{\dagger}$$^{\S}$
\and  and S.\,I.\,Vasiliev$^{\dagger}$$^{\S}$}
\address{$^{\dagger}$ Joint Institute for Nuclear Research,
141980 Dubna, Moscow Region, Russia\\
$^{\ddagger}$Institute of Physics and Technology Problems\\
P.O.\,Box 39, 141980
Dubna, Moscow Region, Russia\\
$^{\S}$Institute for Nuclear Research of Russian Academy of Sciences,
117312 Moscow, Russia}
\date{January 13, 2000}
\maketitle

\bigskip

\begin{abstract}

Experimental search for molecular-nuclear transitions ${\rm 
H}_2{\rm O}$ $\rightarrow$ ${}^{18}{\rm Ne}^*\,(4.522,1^-)$  
$\rightarrow{}^{18}{\rm F}$ ${\rightarrow}$ ${}^{18}{\rm O}$ 
in water molecules was carried out. The measurements were 
performed in a low-background laboratory at the Baksan Neutrino 
Observatory.  Under the assumption that the above transitions 
take place, the estimate for the half-life time of water 
molecule was found to be about $10^{18}$ years.

\end{abstract}

\bigskip

\section{Introduction}

Phenomenon of a long-range effective interaction in some quantum
systems including constituents which are only interacting at
short distances is known already for a long time.  Here, by a
short-range interaction we understand any interaction that falls
off at large distances between particles, i.\,e. as
$r\longrightarrow\infty$, not slower than exponentially. In the context
below, longe-range forces decrease at infinity as some
inverse powers of $r$.

The most pronounced manifestation of this phenomenon appears in
the so-called ``Efimov effect''~\cite{Efimov}. The Efimov effect
consists, in particular, in arising an effective interaction
between a particle and a coupled pair of particles which behaves
as $1/r^2$.  A sufficient condition for this effect to take
place in a three-bosonic system is a closeness of the binding
energies at least for two of pair subsystems to zero. Just
the very small binding energies force the two-body wave function
to be extremely extended generating an effective long-range
interaction of the coupled pair with a complementary particle.

Another example of a long-range effect of short-range
interaction was found by Ya.\,B.\,Zeldovich~\cite{Zeld}. He
showed that if two particles interact via  two potentials, a
short-range one and an arbitrary long-range (say, Coulomb) one,
then the spectrum of the two-body system can be drastically
changed as compared to the case of a long-range
potential only. This change takes place if the short-range
potential generates a bound or resonant state with an energy
sufficiently close to zero.

Thus, in both the above cases, long-range effects take place due
to the fact that a short-range interaction causes a two-body system
to be, nevertheless, very extended.  Having this in mind, one can expect
some enhancement for the probability of transition of the system
from a molecular state to an extended (resonance) nuclear state
as compared to a similar transition into a localized nuclear 
state.  Indeed, model calculations~\cite{BMS} of the overlap 
integral between wave functions of the H$_2$O ($1^-$) molecule 
and a resonant state ($4.522,\,1^-$) of the nucleus $^{18}$Ne 
show an enhancement of this sort.

The purpose of the present experiment is to estimate the life time
of water molecules with respect to the following decay chain
$$
[{\rm H}_2{\rm O}]_{1^-}\longrightarrow
{}^{18}{\rm Ne}^*\,(4.522,1^-)
\longrightarrow{}^{18}{\rm Ne}\,({\rm g.\,s.})
\longrightarrow{}^{18}{\rm F}
{\longrightarrow}{}^{18}{\rm O}+\beta^+.
$$

\section{Experimental approach and results}

A number of examples of nuclear systems with near-threshold
resonances were analyzed from this point of view, among them
are (p,\,p\,,$^{16}$O) and (p,\,$^{17}$O) \cite{Aj,ENDF},
i.\,e. the nuclear constituents of the usual water molecule
H$_{2}$O and hydroxyl ions OH based on the rare oxygen isotope
$^{17}$O (see Table \ref{tableExamples}).

\begin{table}
\caption
{Two examples of nuclear systems with near-threshold resonances}
\label{tableExamples}
\bt{ccccccc}
 & \multicolumn{2}{c}{Few-Atomic System}
 & \multicolumn{4}{c}{Composed Nuclear System (CNS)}\\
\hline
\bt{c} Molecular \\
       System \\
\et
&
\bt{c} Nuclear \\
Subsystem (NS)\\
\et
&
\bt{c}Energy of NS \\
over g.s. of CNS (MeV)\\
\et
&
Nucleus
&
$J^\pi$&E(MeV)&$\Gamma$(keV) \\
\hline
H$_2$O & $^{16}$O$+p+p$ &4.522& $^{18}$Ne&$1^-$&4.519& $<9$ \\
OH$^-$ & $^{17}$O$+p$   &5.607& $^{18}$F &$1^-$&5.604& $<1.2$ \\
\et
\end{table}

\begin{figure}[ht]
\centering
\epsfig{file=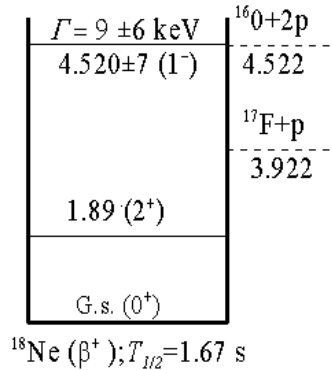,height=6cm}
\caption{Fragment of the nucleus ${}^{18}$Ne spectrum.}
\label{Neon}
\end{figure}

For the fist experimental study in this direction we choose the
system (p,\,p\,,$^{16}$O), i.\,e. the H$_{2}$O molecule. Its
properties, in addition to obvious availability, make this case
the most favorable for the experiment.  From Fig.\,\ref{Neon},
displaying the energy-lever diagram for $^{18}$Ne-nucleus, it is
seen that rotation states 1$^{-}$ of a water molecule and a
highly excited state (1$^{-}$,\,4.522 MeV) of  this nucleus can
be considered as energy degenerate. Thus, a real physical state
with these quantum numbers appears to be a superposition of
molecular and nuclear states. Experimental approach was designed
with taking into account of the properties of both the components of
the superposition pair.

\begin{figure}
\centering
\epsfig{file=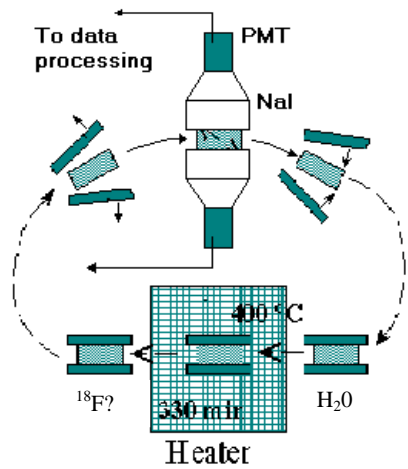,height=10cm}
\vspace{-1cm}
\caption{Scheme of accumulation-measuring procedure.}
\label{cicle}
\end{figure}

As a rule, rotational states are excited only in free molecules
while their population under the conditions of condensed-phase
water is prohibited due to powerful hydrogen bonds \cite{Zaz}.
This  makes carrying out the experiments more difficult, but at
the same time it gives additional opportunities to manipulate
with the hypothetical process of molecular-nuclear transitions in
a water system. So, it becomes possible to use an
accumulation-measuring cycling during searching for resulting
radioactive products. The layout of measurements is shown in
Fig.\,\ref{cicle}.  The accumulation cycle represented heating the water
portion within a sealed out measuring chamber to a critical
point about 647 K, at which a total amount of water
was for sure in a vapor phase regardless of pressure within the
volume. To withstand this rather a high pressure (22.5\,MPa
\cite{Hbook}), the stainless-steel or titanium chambers were
strengthened for the heating period by two thick steel
removable plates at the top and bottom of the chambers. The
latter were thick-wall shortened metallic cylinders, sealed at
the faces by thin membranes almost fully transparent for the
expected annihilation radiation $E_\gamma=511$ keV.

For the measuring period, the plates were removed, and the
chamber was placed after cooling between two
NaI(Tl)-scintillators, operating in the
$\gamma\gamma$-coincidence mode.  First, the measurements were
performed in surface-laboratory conditions at the Institute of
Physics and Technology Problems (Dubna), then the main part of
the experiments was undertaken, this time at a deep
underground laboratory of the Baksan Neutrino Observatory
(Republic of Kabardino-Balkaria, the North Caucasus) of the
Institute for Nuclear Research of the Russian Academy of
Sciences. At the surface-laboratory stage, necessary measurement
procedures were optimized, and a yield of molecular-nuclear
transition in water in the condensed state was estimated.
Analysis of these data showed main sources of the background
counting: true coincidences due to the cosmic muons, decay of
the natural $^{40}$K and daughter products of $^{222}$Rn
($^{214}$Bi and $^{214}$Po). To take into account the background
in the region of ``interest", i.\,e., near $E_\gamma=511$ keV,
some control background measurements were carried out under the
identical conditions (geometry, amount of water, {\it etc.}), in
which heavy water (D$_2$O, 99.0\%-enrichment) was used instead
of the natural one. Within the limits of statistical
fluctuations, both the spectra were identical. For the water
half-life in the condensed-phase state with respect to decay by
the chain H$_2$O$\longrightarrow$$^{18}$Ne($\beta^+$;
1.7\,s)$\longrightarrow{}^{18}$F($\beta^+$; 109 min), a lower limit
was estimated as T$_{1/2}\geq$4.10$^{21}$ years (within the
99\%-confidence level).
At the Baksan Observatory, all the efforts were made to subdue the
background radiation as much as possible. The cosmic muons were
subdued due to power shielding, as the measuring premises were
situated inside a gallery created within the mountain Andyrchi
(one of mountains of the Elbrus environment, where the screening
thickness of a rock achieves about 600\,m of water equivalent.
Walls (the ceiling and floor, as well) were built of a special
uranium-free concrete. The scintillators were prepared on the
basis of materials with a low containment of potassium and
radium. They were, for a long time, located in a deep
underground. Thus all short-lived cosmic-ray-induced activities
should be extinct.  The detection unit was provided with an
additional shielding composed of interchanged layers of pure
tungsten, lead, and copper.

\begin{figure}
\centering
\epsfig{file=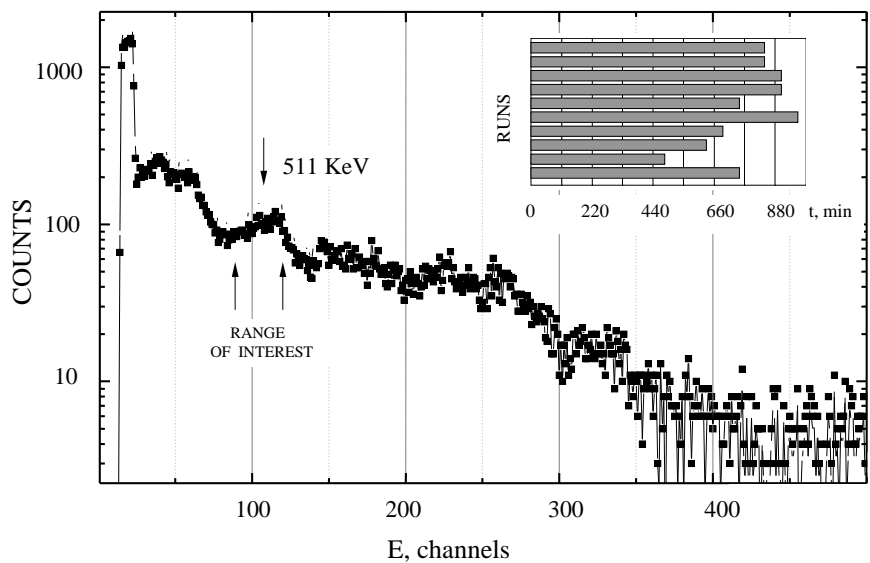,height=10cm}
\caption{The $\gamma\gamma$-coincidence spectrum obtained
during ten measuring cycles. Marked by arrows is
an energy region around $E_\gamma=511$ keV. Time-durations of
measuring periods are displayed in the insertion to the graph.}
\label{spectr_1}
\end{figure}

For every $\gamma$-quantum event, the time was registered with a
channel width 10\,s. This was more than sufficient for a
detailed analysis of the time dynamics in the range of interest
for this experiment. The accumulation interval was chosen to be
about three half-lifes of $^{18}$F, i.\,e. 5.5\,hours. The
measuring time-schedule for these runs is inserted in
Fig.\,\ref{spectr_1}. Intervals of about four half-lives of
$^{18}$F, i.\,e. $\sim440$ min each, were considered as
time-periods when the decay of hypothetically accumulated
$^{18}$F could yet give some contribution to the counting in the
region of $E_\gamma=511$ keV.  The remaining part of measuring
cycles was used to calculate the background counting in the same
energy range.

Preliminary analysis of the results was carried out by means of 
comparing counting rates in two above measurement periods, each 
of a total duration $\sim4000$ min for ten cycles.  A certain 
excess (at a level about 1-2 RMS deviation) of the counting rate 
in the region of $E_\gamma=511$ keV was observed in the first 
period as compared to the second one, i.\,e. to the background.  
We emphasize that the above excess was non-stationary.  Time 
dependence of the non-stationary process approximately 
corresponded to the half-life of $^{18}$F.  Under the assumption 
that the observed non-stationary component of the effect is 
associated with the accumulation and decay of the nuclei 
$^{18}$F, our estimate for the efficient half-life of the water 
molecule with respect to the nuclear channel under consideration 
is $T_{1/2}\sim10^{18}$ years.

Although the total statistics was not sufficient to make a 
decisive conclusion, the results obtained are rather encouraging 
for further experiments in this intriguing direction with new 
more sensitive approaches.  More attention should be paid to 
analysis of population rates of suitable molecular states under 
the experimental conditions.  Also some further theoretical 
consideration are desirable regarding the fusion probability 
estimates for this and other molecular systems. 

\acknowledgements
This work was supported by the Russian Foundation for Basic
Research (grant \#\,98-02-16884).


\begin{thebibliography}{99}

\bibitem{Efimov} V. Efimov, Nucl. Phys. A {\bf 210} (1973), 157.

\bibitem{Zeld} Ya. B. Zeldovich, Physics of the Solid State {\bf
1} (1959), 1637 (Russian).

\bibitem{BMS}  V.B. Belyaev, A.K. Motovilov, W. Sandhas, 
Physics--Doklady {\bf 41} (1996), 514; JINR Rapid Communications 
{\bf 6\,[74]} (1995), 5 (LANL E-print nucl-th/9601021). 

\bibitem{Aj}  F. Ajzenberg-Selove,
Nucl. Phys. A. {\bf 392} (1983), 1.

\bibitem{ENDF} Evaluated Nuclear Structure Data Files (ENSDF),
National Nuclear Data Center (NNDC) at
Brookhaven National Laboratory, http://www.tunl.duke.edu/NuclData.

\bibitem{Zaz}  G.N. Zatsepina, {\it Physical Properties and Structure
of Water,} Moscow University Press, Moscow, 1998  (Russian). P.\,57.

\bibitem{Hbook}  V.V. Ignatiev, V.A. Krivoruchko, A.I. Migachev,
In: {\it Physical Constants, Handbook} (Z.D. Andreenko
et al., ed.). Moscow, Energoatomizdat, 1991  (Russian). P.\,254.


\end{thebibliography}
\end{document}